\documentclass[fleqn,twoside]{article}
\usepackage{espcrc2}

\usepackage{axodraw}
\usepackage{graphicx}
\usepackage{psfig}
\usepackage[figuresright]{rotating}

\newcommand{\AmS}{{\protect\the\textfont2
  A\kern-.1667em\lower.5ex\hbox{M}\kern-.125emS}}

\title{{\tt PHASE} - An event generator for six fermion physics at the LHC}

\author{E. Accomando \address[MCSD]{Dipartimento di Fisica Teorica, 
Universit\`a di Torino,\\
Via P. Giuria 1,  10125, Torino, Italy \\
and \\
INFN, Sezione di Torino,\\
Via P. Giuria 1,  10125, Torino, Italy},
A. Ballestrero \addressmark ,
E. Maina \addressmark}

\begin{document}

\begin{abstract}
{\tt{PHASE}} is a Monte Carlo event generator, under construction, for all 
Standard Model processes with six fermions in the final state at the LHC.
It employs the full set of tree level Feynman diagrams, taking into
account fermion masses for b quarks. 
The program can generate unweighted events for any subset of
all six fermion final states in a single run, by making use of dedicated 
pre-samples. An interface to hadronization is provided.
\vspace{1pc} \end{abstract}

\maketitle

\section{Introduction}

At future colliders, many-particle final states will be accessible 
with much more statistics. Among these, six fermion signals are of 
particular interest for their relevance in top-quark physics, Higgs boson
production in the intermediate mass range, vector boson scattering, and 
quartic gauge boson coupling analyses.

Typically, multi-particle final states can come from the decay of
intermediate unstable particles that are produced as resonances in
subprocesses. Relying on this, the most commonly used generators Pythia
\cite{pythia} and Herwig \cite{herwig} are entirely based on the
{\it production $\times$ decay} approach, i.e. narrow width
(NWA) and effective vector boson approximation (EVBA).
While the general expectation is that NWA
(with possible improvements by spin correlation between production and
decay or by Breit-Wigner convolution) can give an order ten percent unaccuracy,
the EVBA is surely more problematic.
In the basic leading-log formulation, it does not allow to freely implement 
any kind of kinematical cuts. The parton coming out after the vector boson
emission is in fact produced in a fully inclusive way. 
Even in its improved realization, which accounts for the exact phase space,
the EVBA cannot always reproduce the correct kinematics and event rate.
The result strongly depends on the applied cuts, and there is no unique
recipe. Therefore, its reliability must be always cross-checked against a
complete program implementing exact calculations.

A further reason for going towards full computations is connected to 
gauge invariance and the appearance of strong and extremely delicate 
cancellations 
at high energy among Feynman diagrams contributing to the same final state.
These interferences can take place both within the signal diagram set and
between signal and irreducible background diagrams.
Thus exact matrix elements must be computed. This means evaluating thousands
of diagrams for a huge variety of possible processes.
%But this is just one of the stages of this complex ground. In fact, owing
%to their increased structure, the amplitudes show a very complicated
%resonant structure which must be integrated over the full phase space. A
%good integration is intimately connected to a good generation efficiency,
%which is a key parameter. So, new integration strategies like
%multi-channel and related variations have been developed.

In the last decade, a big effort has been dedicated to the implementation
of event generators for multi-parton production. At present, the following 
codes are available: {\tt Amegic} \cite{Amegic}, {\tt CompHEP} 
\cite{Comphep}, {\tt Grace} \cite{Grace}, {\tt Madgraph} \cite{Madgraph}, 
{\tt Phegas \& Helac} \cite{Phegas}, {\tt O'Mega $\&$ Whizard} \cite{Omega}.
These are {\it multi-purpose} generators, which in principle can compute any 
tree level process. 
A different approach is given by event generators {\it dedicated} to specific 
classes and topologies of final states.  
A recent example of this kind of generator for LHC physics is {\tt Alpgen} 
\cite{Alpgen}.

\section{{\tt PHASE}: PHact Adaptive Six-fermion Event generator}

In this section, we present the general features of the Monte Carlo
event generator {\tt PHASE} \cite{PHASE}, which 
is fully {\it{dedicated}} to six-fermion physics. In its first version, the
PHASE project is designed for all Standard Model processes $pp\rightarrow
6f$ in lowest order.

\begin{table*}[htb]
\caption{Processes and groups}
\label{table:1}
\newcommand{\m}{\hphantom{$-$}}
\newcommand{\cc}[1]{\multicolumn{1}{c}{#1}}
\renewcommand{\tabcolsep}{2pc} % enlarge column spacing
\renewcommand{\arraystretch}{1.2} % enlarge line spacing
\begin{tabular}{@{}llll}
\hline
particles & type & diagrams & process number \\
\hline
$c\bar s d\bar u c\bar s l\bar\nu$ & 4W & 202={\bf{101}}$\times$2 & 6+2\\
$u\bar u u\bar u c\bar s l\bar\nu$ & 2Z2W & 422={\bf{211}}$\times$2 & 6+2\\
$u\bar u c\bar c c\bar s l\bar\nu$ & 2Z2W & 422={\bf{211}}$\times$2 & 10+1 \\
$u\bar u s\bar s c\bar s l\bar\nu$ & 2Z2W & 422={\bf{211}}$\times$2 & 10+1 \\
$u\bar u b\bar b c\bar s l\bar\nu$ & 2Z2W & 233={\bf{211}}+22 & 15+0 \\
$d\bar d d\bar d c\bar s l\bar\nu$ & 2Z2W & 422={\bf{211}}$\times$2 & 6+2\\
$d\bar d c\bar c c\bar s l\bar\nu$ & 2Z2W & 422={\bf{211}}$\times$2 & 10+1 \\
$d\bar d s\bar s c\bar s l\bar\nu$ & 2Z2W & 422={\bf{211}}$\times$2 & 10+1 \\
$d\bar d b\bar b c\bar s l\bar\nu$ & 2Z2W & 233={\bf{211}}+22 & 15+0 \\
$c\bar c c\bar c c\bar s l\bar\nu$ & 2Z2W & 1266={\bf{211}}$\times$6 & 3+2 \\
$c\bar c b\bar b c\bar s l\bar\nu$ & 2Z2W &
466=({\bf{211}}+{\bf{22}})$\times$2 & 10+1 \\
$s\bar s s\bar s c\bar s l\bar\nu$ & 2Z2W & 1266={\bf{211}}$\times$6 & 3+2 \\
$s\bar s b\bar b c\bar s l\bar\nu$ & 2Z2W &
466=({\bf{211}}+{\bf{22}})$\times$2 & 10+1 \\
$b\bar b b\bar b c\bar s l\bar\nu$ & 2Z2W &
610=({\bf{211}}+{\bf{94}})$\times$2 & 6+2 \\
$u\bar u d\bar d c\bar s l\bar\nu$ & 2Z2W+4W & 312={\bf{101}}+{\bf{211}} &
15+0\\
$c\bar c s\bar s c\bar s l\bar\nu$ & 2Z2W+4W &
1046={\bf{101}}$\times$2+{\bf{211}}$\times$4 & 6+2 \\
%TOTAL &  &  & 141+20 \\
\hline
\end{tabular}\\[2pt]
\end{table*}

\subsection{Processes}

At the present stage, {\tt PHASE} includes $O(\alpha^6)$ electroweak
processes with a neutrino in the final state, $pp\rightarrow 4ql\nu_l$.
By making use of symmetries, all these channels can be classified into
16 groups which are enumerated in Table 1.
By selecting two initial quarks in each particle group, one can obtain all
possible processes whose number is given in the last entry.
Taking into account charge conjugation and family symmetry, one ends up
with more than one thousand processes. In some of them, fermions can only 
be paired into charged currents (4W), in others into two charged and two 
neutral currents (2Z2W) or into a mixed combination (2Z2W+4W). 

It should be noticed 
that the amplitudes of the above mentioned 16 groups are not all
independent. They are in fact combinations of only 4 basic sets of Feynman
diagrams (101, 211, 22, 94). This means that all thousand processes can be 
implemented using just few building blocks. The immediate advantage is
that any modification, like including new couplings or vertices, has to be
done only in a very restricted area of the program and then
it will be automatically communicated to all processes.

\subsection{Helicity amplitudes}

{\tt PHASE} works with exact matrix elements, thus providing a complete
description of signal and irreducible background. All amplitudes are
written with the help of the program PHACT \cite{PHACT}, which
is based on the helicity formalism of ref. \cite{helamp}.
This method allows one to calculate parts of diagrams of increasing
size and complexity, and store them for later use. In this way, common
subdiagrams are evaluated just once, with a substantial efficiency gain.
The formalism is appropriate both for massless and massive fermions. In 
{\tt PHASE}, fermion masses are exactly taken into account for {\it b}
quarks. 

\subsection{Phase space integration}

Since a single process can contain hundreds of diagrams, the amplitude 
peaking structure is generally rather complex. As a consequence the 
15-dimensional phase space has untrivial kinematical regions corresponding
to the matrix element singularities. In order to gain in accuracy and 
efficiency, {\tt PHASE} relies on a new integration method, which combines
together the adaptivity principle \'a la {\tt VEGAS} \cite{vegas}, and the
multi-channel strategy.
The outcome is that {\tt PHASE} adapts to different kinematical cuts and 
resulting peaks with good efficiency. In contrast to the pure multi-channel
approach, where one has to introduce a phase space parametrization with
appropriate mappings, called channel, for each propagator appearing in the
amplitude, here a maximum of four channels is required. Moreover, owing
again to adaptivity, only a rough estimate of the relative weights
of the channels is sufficient for an accurate integration.
The drawback is that each channel must be integrated separately.

During the integration of the single process, integration grids will be 
generated (one for each channel). These grids optimize the integration
itself and constitute the basic ingredient for the so called {\it
one-shot} event generation we are going to describe in the next section. 
The important feature of the integration grids is that they are computed
just
once, with the loosest set of cuts in order to retain as much information 
as possible, and stored for later use.

\subsection{{\it One-shot} event generation}

After generating the integration grids, the {\it one-shot} procedure can
start.
This is one of the main features of {\tt PHASE}. In fact, it allows the user to
generate unweighted events not on a process by process basis but for any 
possible set of processes in just a single run, giving at the end a complete 
event sample where all included final states appear in the right relative 
proportion.

The general method for the {\it one-shot} generation is very similar to
the one used in {\tt WPHACT} and described in ref.\cite{wphact}, but
generalized to six-fermion processes.
Given the integration grids described in the previous section, {\tt PHASE} 
will read from these files all necessary information and build up for every 
single channel its probability and a maximum normalized to it. According to 
this probability, one channel will be extracted at a time, and an event will 
be generated with a frequency determined by its 
grid. The event will be then compared with the normalized maximum in order to 
keep or reject it. The procedure is repeated until the required number of 
unweighted events is produced.

The generated events can then be passed to Pythia, in order to simulate 
observable
final states via showering and hadronization. In this way, one can have
a complete and accurate tool for realistic experimental simulations.
This step is performed according to the "Les Houches accord" 
\cite{leshouches}, a set of common blocks for passing event configurations 
from parton level generators to parton shower and hadronization packages.

\begin{table*}[htb]
\caption{Gauge invariance and cancellations.}
\label{table:2}
\newcommand{\m}{\hphantom{$-$}}
\newcommand{\cc}[1]{\multicolumn{1}{c}{#1}}
\renewcommand{\tabcolsep}{2pc} % enlarge column spacing
\renewcommand{\arraystretch}{1.2} % enlarge line spacing
\begin{tabular}{@{}lllll}
\hline
$M_h$  & signal (pb)  & total cross section (pb)  \\
\hline
120 GeV &  0.2106672   &   0.1319138E-02 \\
200 GeV &  0.2174115   &   0.7960299E-02 \\
500 GeV &  0.2114054   &   0.2804993E-02 \\
2000 GeV &  0.2100076   &   0.1490797E-02 \\
No Higgs &  0.2099891   &    0.1468983E-02 \\
\hline
\end{tabular}\\[2pt]
\end{table*}

\section{Results for Vector boson scattering}

We consider a typical channel including the subprocess
$WW\rightarrow WW$ at the LHC. We analyze the process
$ud\rightarrow udc\bar s\mu\bar\nu$ in the region where we have
the distinctive two forward jets signature. To this aim, we apply the
following cuts:
\begin{center}
$E_T(u,d)\ge$ 20 GeV~~~~~~~$P_T(u,d)\ge$ 10 GeV

-5.5$\le\eta (u)\le$-1~~~~~~~1$\le\eta (d)\le$5.5

-10 GeV $\le M(c\bar s, l\bar\nu )-M_W\le$ 10 GeV
\end{center}
The first effect we investigate is connected to gauge invariance. 
It was already pointed out by Kleiss and Stirling in an old
paper \cite{kleiss} that selecting only the subset of vector boson signal
diagrams from the full amplitude could give meaningless results. This
argument relies on the fact that the delicate gauge cancellations get
destroyed by the off-shellness of the two initial vector bosons.   
We have checked this issue in more detail. In Table 2, we present the
total cross section and the pure vector boson scattering signal for
different Higgs masses, in unitary gauge. As one can see, the full
cross section is two
order of magnitude smaller than the signal. This poses an important
question on what kind of signal definition one can give at high energy.
This issue should receive more attention.
In Fig.1 we show the WW invariant mass distribution for different helicity
configurations of the two reconstructed W's.
Starting from the left, from top to bottom we plot the UU, TT, TL and LL
contributions, where U, T and L are for unpolarized, transverse and 
longitudinal gauge boson. Additional $P_T(c\bar s,l\bar\nu )\ge M_W$, 
$E_T(c,\bar s,l)\ge$20 GeV and $P_T(c,\bar s,l)\ge$ 10 GeV cuts are included. 

  \unitlength 1cm
  \begin{picture}(15.,8.5)
%  \caption{Figure 1. WW invariant mass distribution.}
  \put(0.,9.){\includegraphics{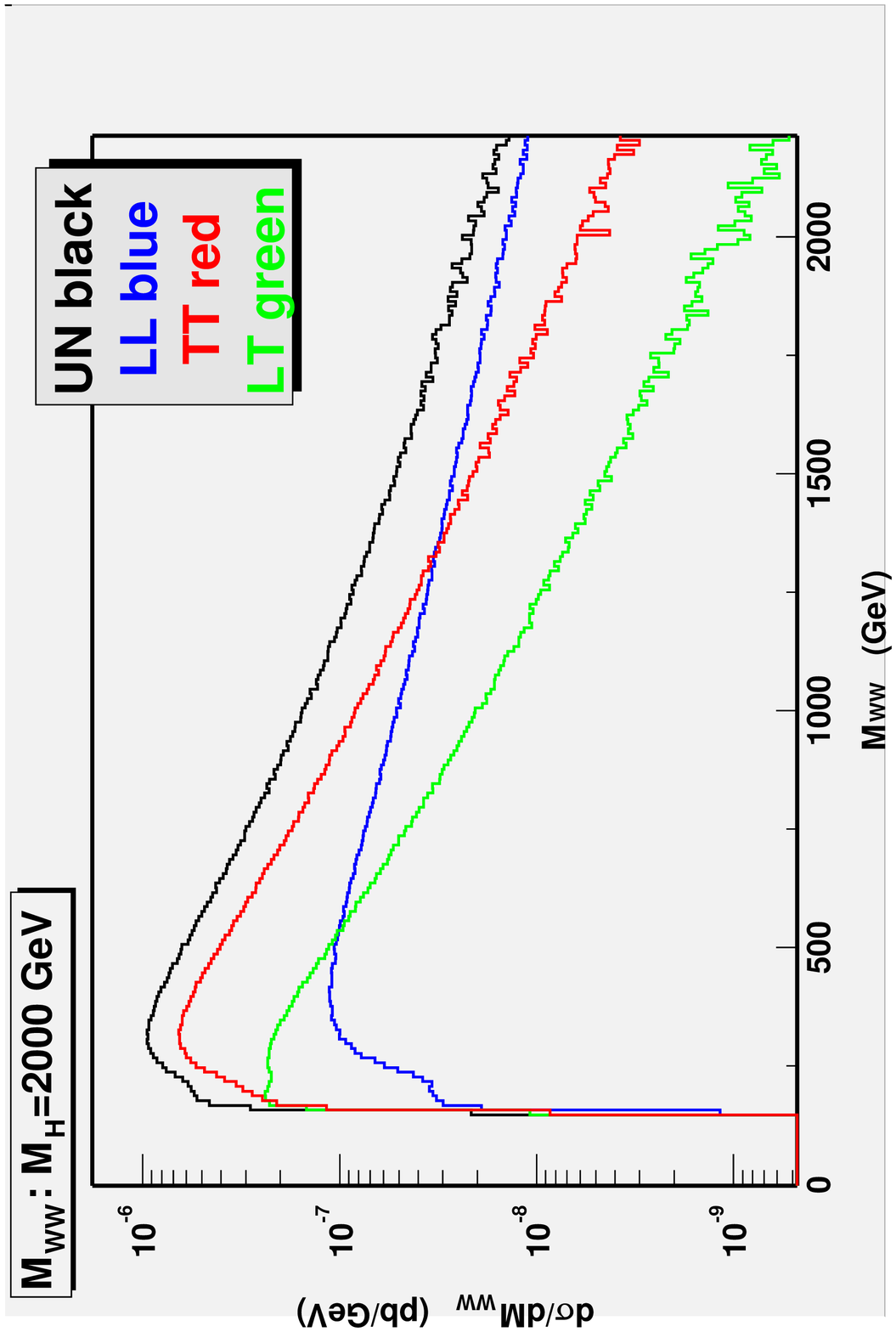}}
  \end{picture}
\vskip -1.cm
\centerline{Figure 1. WW invariant mass distribution.}

\end{document}